\documentclass[sigconf,natbib=true]{acmart}

\AtBeginDocument{%
  \providecommand\BibTeX{{%
    \normalfont B\kern-0.5em{\scshape i\kern-0.25em b}\kern-0.8em\TeX}}}

\begin{document}

\title{MCMF: Multi-Constraints With Merging Features Bid Optimization in Online Display Advertising}

\author{%
Xiao Wang, Shaoguo Liu*, Yidong Jia, Yuxin Fu, Yufang Yu, Liang Wang, Bo Zheng
}
\affiliation{%
  \institution{Alibaba Group}
  \city{Beijing}
  \country{China}}
\email{
  {ranxiang.wx, shaoguo.lsg, yidong.jyd, yuxin.fyx,
  yuyufang.yyf, liangbo.wl, bozheng}@alibaba-inc.com
}
\thanks{*Corresponding Author}

\renewcommand{\shortauthors}{Wang and Liu, et al.}

\begin{abstract}
  In the Real-Time Bidding (RTB), advertisers are increasingly relying on bid optimization to gain more conversions (i.e trade or arrival).
  Currently, the efficiency of bid optimization is still challenged by the \textbf{(1) sparse feedback}, \textbf{(2) the budget management separated from the optimization}, and \textbf{(3) absence of bidding environment modeling}.
  The conversion feedback is delayed and sparse, yet most methods rely on dense input (impression or click).
  Furthermore, most approaches are implemented in two stages: optimum formulation and budget management, but the separation always degrades performance.
  Meanwhile, absence of bidding environment modeling, model-free controllers are commonly utilized, which perform poorly on sparse feedback and lead to control instability.
  
  We address these challenges and provide the Multi-Constraints with Merging Features (MCMF) framework.
  It collects various bidding statuses as merging features to promise performance on the sparse and delayed feedback.
  A cost function is formulated as dynamic optimum solution with budget management, the optimization and budget management are not separated.
  According to the cost function, the approximated gradients based on the Hebbian Learning Rule are capable of updating the MCMF, even without modeling of the bidding environment.
  Our technique performs the best in the open dataset and provides stable budget management even in extreme sparsity.
  The MCMF is applied in our real RTB production and we get 2.69\% more conversions with 2.46\% fewer expenditures.
\end{abstract}

\setcopyright{none}
\settopmatter{printacmref=false,printfolios=true} 
\renewcommand\footnotetextcopyrightpermission[1]{} 

\maketitle
\section{Introduction and Related Works}
\begin{figure}[t]
    \vspace{-0.2cm}
    \setlength{\abovedisplayskip}{0cm}
    \setlength{\belowdisplayskip}{-0.5cm}
    \setlength{\abovecaptionskip}{0cm}
    \setlength{\belowcaptionskip}{-0.6cm}
    \centering
    \includegraphics[width=0.85\linewidth]{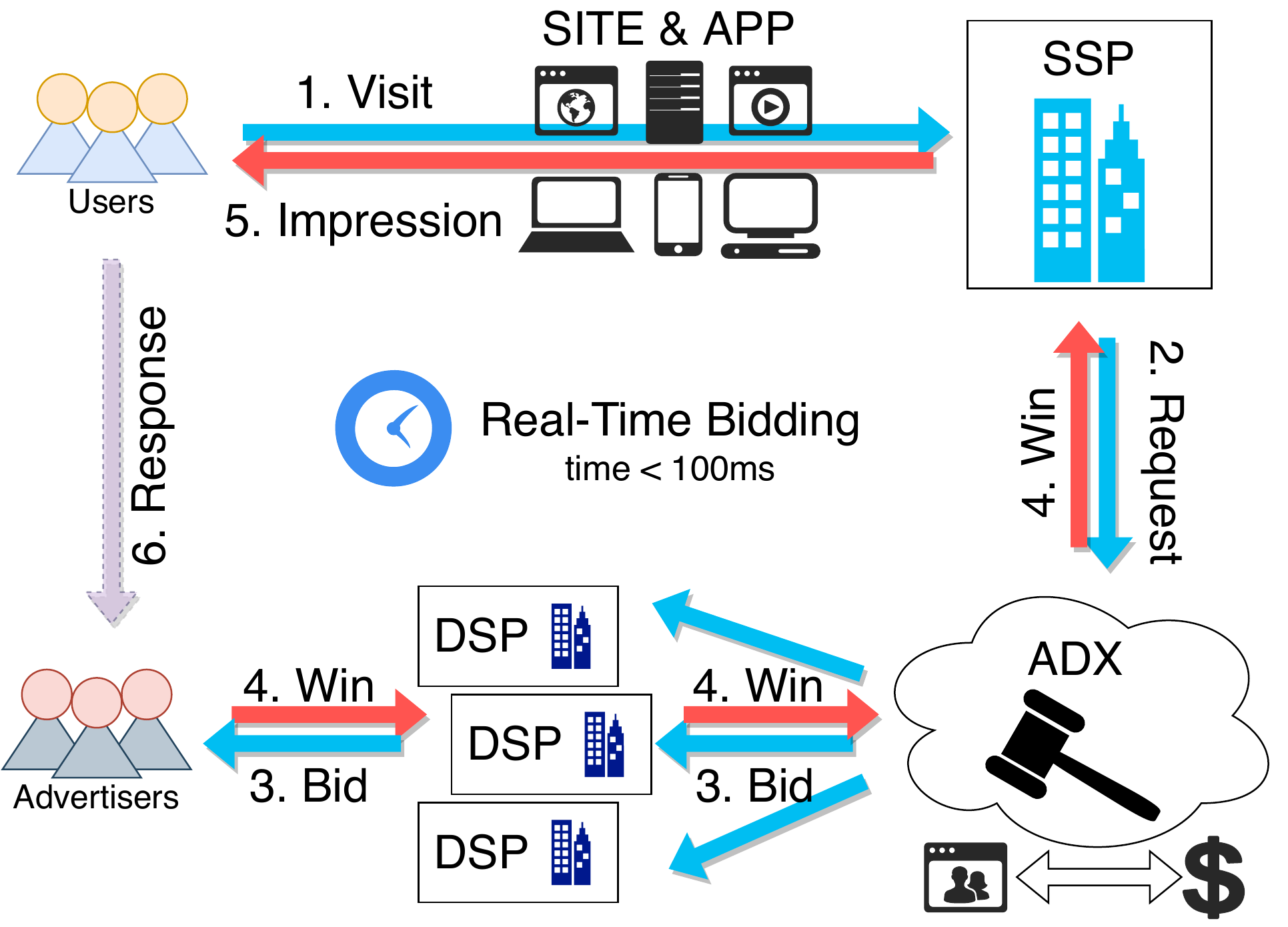}
    \caption{The Real-Time Bidding.}
    \label{fig:rtb_processing}
\end{figure}

The Real-Time Bidding (RTB), as shown in Figure \ref{fig:rtb_processing}, can manage each ad impression through real-time auctions.
Publishers offer advertising impressions to advertisers by organizing a fair and transparent real-time auction.
When a user \textbf{visits(1)} a website or mobile app page, the Supply-Side Platform (SSP) sends and broadcasts a \textbf{request(2)} of an ad display opportunity to the Demand-Side Platforms (DSPs) via the ad exchange (ADX).
An auction is held on the ADX in a very short time frame, and each DSP decides on a final \textbf{bid(3)} price for this request.
The highest bidder \textbf{wins(4)} the impression opportunity, and the actual payment (cost) could be sent to the win DSP from SSP\cite{yuan2014survey}.
Following the \textbf{impression(5)}, the user's \textbf{response(6)} will be sent to the DSP as feedback.
The auction occurs in the split second to load a Web page.
The users' responses include the impression, click, order, or pay.
From the impression to pay, the delay and sparsity are increased sharply.

The first challenge to bid optimization is the transition from maximal dense feedback to sparse feedback.
The bid optimization decides a final bid price to win more opportunities and gain more conversions.
It is often treated as an optimization problem for maximizing the bidder's utility with multi-constraints.
These constraints are so-called Key Performance Indicators (KPIs)\cite{kitts2017ad,tashman2020dynamic} as expected Cost-Per-Click (eCPC), expected Cost-Per-Milles (eCPM) or budget limitation.
However, the formulating of these optimizations only takes those dense and immediate feedback (impression or click) as the bidder's utility.
Currently, advertisers (bidders) are willing to pay for the performance of the conversion, which always feeds back sparsely with a long delay.
The traditional method does not adapt to the sparse feedback and leads to the spending out of control.
The spending would exceed the daily budget in a very short time.
Attaching pacing mechanisms \cite{xu2015smart,lin2020budget} can relieve the problem, but it only put off the time of exceeding budget.
The spending out of control still degrades the performance of bid optimization.
\begin{figure*}[ht]
    \vspace{-0.5cm}
    \setlength{\abovedisplayskip}{0cm}
    \setlength{\belowdisplayskip}{-0.5cm}
    \setlength{\abovecaptionskip}{0cm}
    \setlength{\belowcaptionskip}{-0.5cm}
    \centering
    \includegraphics[width=0.95\linewidth]{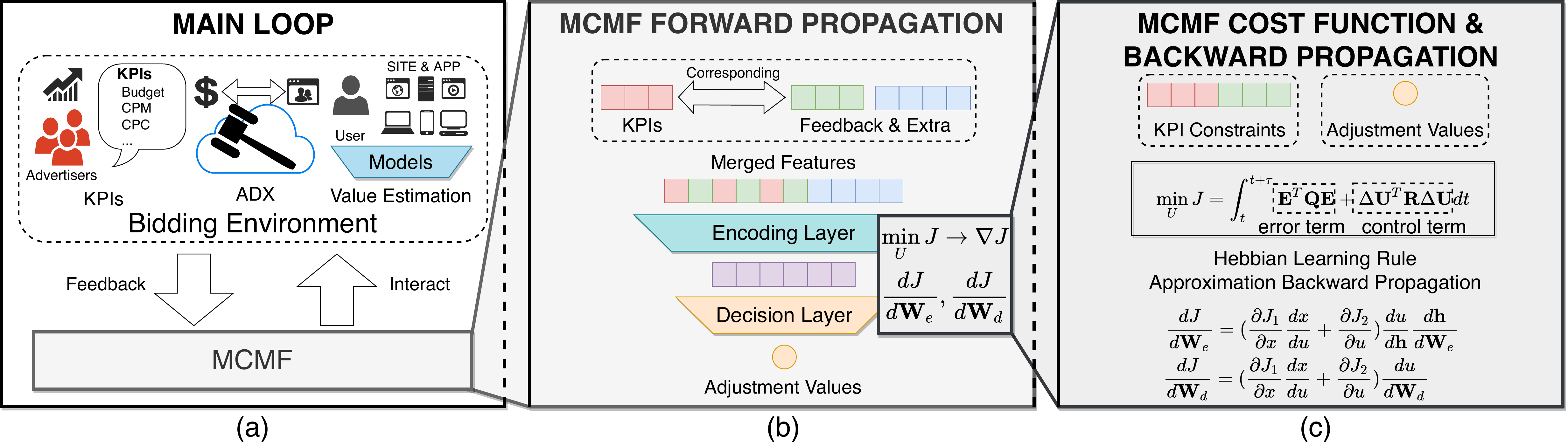}
    \caption{The MCMF framework.(a)main loop;(b)forward propagation;(c)cost function and backward propagation.}
    \label{fig:mcmf_full}
\end{figure*}

Traditional bid optimization suffers from the separation of optimum formulation and budget management, which is the second challenge of bid optimization.
Most approaches formulate an optimal solution based on the linear programming prime-dual formulation to maximum the bidder's utility\cite{yuan2014survey,tashman2020dynamic,liu2020effective,lin2020budget}.
The advertiser's KPIs are treated as the constraints of the optimization.
The prime-dual formulation can derive a controllable function and employ strategies\cite{liu2020effective} or controllers\cite{perlich2012bid} to adjust the bid price.
When the solution's constraints are changed or reconstructed, the prime-dual formulation and the derived function should be reformulated.
So the solution is static and not flexible enough to deal with the dynamic bidding environment.
Alternatively, by manipulating agents, Reinforcement Learning-based approaches can yield a dynamic optimal solution.
Agents are trained to change bid prices based on policy rewards\cite{yang2020motiac,lu2019reinforcement,cai2017real}.
Training the agents requires a large amount of actions and conversion rewards data, but the sparsity of the feedback may result in less-than-optimal results.

The absence of bidding environment modeling is the third challenge of bid optimization.
The very dynamic and variable bidding environment can hardly be represented or modeled exactly.
Researches are proposed with very complicated models and plenty of parameters to represent the bidding environment\cite{he2019identification,karlsson2020feedback}.
Even though, the model is not so exact to every bidding environment.
Furthermore, lacking of the exact model, the advanced optimal controllers can hardly be utilized in bid optimization.
The model free controllers such as the PID (Proportion Integration Differentiation) are commonly used\cite{yang2019bid}.
The PID controller is not an ideal optimum controller, it is sensitive to its hyper-parameters and is frequently unstable due to latency.
Although the optimal solution is right, the PID controller may provide unexpected consequences. 

In this paper, we propose a unified framework named Multi-Constraints with Merging Features (MCMF) for bid optimization under sparse feedback even without an exact model of a dynamic bidding environment.
Our MCMF is a gradient-based two-layers Multi Layer Perception (MLP), which updates weights by following the Hebbian Learning Rule\cite{munakata2004hebbian}.
\textbf{(1) It collects various bidding statuses as merging features to forward propagation and decide a bid price.
This can promise the bidding performance on the condition of sparse and delay feedback.
(2) The weights of MCMF are updated in backward propagation by minimizing the cost function.
The designed cost function can maximize the bidder's utility dynamically with stabilized budget spending.
Due to the design, the utility maximization and budget management are uniformed into one cost function instead of optimization in separate stages.
(3) We propose approximated gradients for the backward propagation by following the Hebbian Learning Rule.
This can ensure that the gradient updating still works in each time period even in absence of exact bidding environment model.}
Our MCMF performs the best in a popular open dataset and provides a stable budget management even the feedback is extremely sparse.
We also applied our method in our RTB production and get a well performance in conversion.
\section{Problem formulation}
The feedback from the bidding environment involves various statuses from the ADX, advertisers' KPIs, and value estimation during the main loop in Figure \ref{fig:mcmf_full}(a).
MCMF is a two-layer multi-layer perception (MLP). As shown in Figure \ref{fig:mcmf_full}(b), the input features are merged by various bidding statuses
The status includes the KPIs with the corresponding feedback and extra features.
The merged features can promise the performance of MCMF under the condition of sparse and delayed feedback.
We designed a cost function to balance the bidder's utility maximization and budget management dynamically as Figure \ref{fig:mcmf_full}(c).
In every time period, the MLP's weights are updated by minimizing the cost function.
The updating is an approximated backward propagation by following the Hebbian Learning Rule.
Even with no exact modelling of bidding environment, the updating is still working.
We formulate the MCMF in Section\ref{sec:bid_adjustment}, define the cost function in Section \ref{sec:cost_func}, and propose the approximated backward propagation in Section \ref{sec:backward}.
\subsection{Bid adjustment}
\label{sec:bid_adjustment}
In the main loop, we defined a bid adjustment value $u^{(t)}$ in current time period $t$.
The adjusted $eCPM$ (estimated cost-per-milles) can be denoted by the product of the \textit{original bid price} and the adjustment value $u^{(t)}$ in Eq.\ref{eq:bid_adjust_value}.
The original bid price is a production of predicted click-through rate $pCTR$, predicted conversion rate $pCVR$ and the expected pay-per-conversion (PPC) which is denoted by $PPC_{e}$.
\begin{equation}
    \label{eq:bid_adjust_value}
    eCPM = \underbrace{1000 \cdot pCTR \cdot pCVR \cdot PPC_{e}}_{original \, bid \, price} \cdot u^{(t)} \\ 
\end{equation}

The adjustment value $u^{(t)}$ can be calculated by a two-layers MLP forward propagation in Eq.\ref{eq:layers}.
The $\mathbf{W}_e$ and $\mathbf{W}_d$ are denoted the weights of \textbf{Encoding Layer} and \textbf{Decision layer}.
We marked the encoded vector $\mathbf{h}$ for explaining backward propagation in section \ref{sec:backward}.
The input vector $\mathbf{x} = \left[z_1, x_1, \dots, z_i, x_i, \mathbf{v} \right]$ can be concatenated by the set KPI $z_i$ with the corresponding feedback $x_i$ and other accumulated feedback merged feature $\mathbf{v}$.
\begin{equation}
    \label{eq:layers}
    \mathbf{h} = \mathbf{W}_e \cdot \mathbf{x}, \quad\quad
    \mathbf{U} = \sigma(\mathbf{W}_d \cdot \mathbf{h})
\end{equation}
\subsection{Cost function}
\label{sec:cost_func}
We define a cost function that is capable of optimizing dynamically and it uniforms utility maximization and budget management.
The cost function is qudratic with a \textit{KPIs error term} and a \textit{control term}.
A KPI error vector $\mathbf{E}$ and an error-parameter matrix $\mathbf{Q}$ are used to calculate the KPI \textit{error term}, which can balance each KPI's following error.
The adjustment error vector $\Delta U$ and control-parameter matrix $\mathbf{R}$ are used to calculate the \textit{control term}, which represents the control cost and stability of the previous two periods.
In Eq. \ref{eq:target_error}, the error vector $\mathbf{E}$ is the difference between KPIs (e.g., expected PPC or budget) $z_i$ and the corresponding accumulated feedback $\sum_{t}{x_i}$.
In Eq. \ref{eq:delta}, the adjustment error vector $\Delta\mathbf{U}$ is the difference of two adjacent time period adjustment values.
For a more smooth control, we recommend using the last $\tau$ time period accumulation loss for backward propagation.
\begin{equation}
    \label{eq:cost}
    J = \int_t^{t+\tau}\underbrace{\mathbf{E}^T\mathbf{Q}\mathbf{E}}_{error \, term} + \underbrace{\Delta\mathbf{U}^T\mathbf{R}\Delta\mathbf{U}}_{control \, term}dt
\end{equation}
\begin{equation}
    \label{eq:target_error}
    \mathbf{E}  = [
    \begin{array}{cccc}
    \sum\limits_{t}{x_1^{(t)}} - z_1, & \sum\limits_{t}{x_2^{(t)}} - z_2, & \dots, & \sum\limits_{t}{x_i^{(t)}} - z_i \\
    \end{array}
    ]^T
\end{equation}
\begin{equation}
    \setlength{\abovedisplayskip}{3pt}
    \setlength{\belowdisplayskip}{3pt}
    \label{eq:delta}
    \Delta\mathbf{U}^{(t)} = \mathbf{U}^{(t)} - \mathbf{U}^{(t-1)} = 
    [
        u_1^{(t)} - u_1^{(t-1)}, \dots, u_j^{(t)} - u_j^{(t-1)}
    ]^T
\end{equation}

The error-parameter matrix is $\mathbf{Q}=diag\{q_1,\dots,q_i\}$, $q_i>0$, and the control-parameter matrix is $\mathbf{R}=diag\{r_1, \dots, r_j\}$, $r_j>0$.
When $||\mathbf{Q}||>>||\mathbf{R}||$, the next time period adjustment value can be determined to bring the most recent feedback closer to the target KPIs.
On the contrary, when $||\mathbf{R}||>>||\mathbf{Q}||$, the determined adjustment value is changed smoothly from the last time period.
\subsection{Approximated gradients update}
\label{sec:backward}
According to the chain rule, the gradients for backward propagation should be calculated by minimizing the cost function $J$.
The gradients from the ADX cannot be computed since no model can represent the ADX.
We propose an approximated gradients for the backward propagation by referring to the Hebbian Learning Rule\cite{munakata2004hebbian} in Eq.\ref{eq:grad1} and Eq.\ref{eq:grad2}.
\begin{equation}
    \label{eq:grad1}
    \frac{dJ}{d\mathbf{W}_e} = (\frac{\partial J_1}{\partial x}\frac{dx}{du} + \frac{\partial J_2}{\partial u})\frac{du}{d\mathbf{h}}\frac{d\mathbf{h}}{d\mathbf{W}_e}
\end{equation}
\begin{equation}
    \label{eq:grad2}
    \frac{dJ}{d\mathbf{W}_d} = (\frac{\partial J_1}{\partial x}\frac{dx}{du} + \frac{\partial J_2}{\partial u})\frac{du}{d\mathbf{W}_d}
\end{equation}

The partial gradients $\frac{\partial J_1}{\partial x}$ and $\frac{\partial J_2}{\partial u}$ are denoted by Eq.\ref{eq:partial_j1} and Eq.\ref{eq:partial_j2}.
In Eq.\ref{eq:partial_j2}, the $(t-1)$-st adjustment value $u^{(t-1)}$ should be treated as a constant value in the latest time period.
\begin{equation}
    \label{eq:partial_j1}
    \frac{\partial J_1}{\partial x} = 2q(x^{(t)}-z)
\end{equation}
\begin{equation}
    \label{eq:partial_j2}
    \frac{\partial J_2}{\partial u} = 2r(u^{(t)} - u^{(t-1)}) 
\end{equation}

With no exact ADX model, we employ a sign function to approximate the gradients of $\frac{dx}{du}$ in Eq.\ref{eq:dxdu}, which can obtain similar trends to the actual gradients of the ADX function.
We use $x = f_{ADX}(u)$ to represent the input and output of the ADX for each time period.
\begin{equation}
    \label{eq:dxdu}
    \frac{dx}{du} = \frac{df_{ADX}}{du} \approx sign((x^{(t)} - x^{(t-1)})\cdot (u^{(t)}-u^{(t-1)}))
\end{equation}

The output $u^{(t)}$ is yielded from the Sigmoid function.
The derivative of the Sigmoid function is included in gradients of $\mathbf{h}$ and $\mathbf{W}_d$, which are denoted as Eq.\ref{eq:grad_sigma_wd}. 
\begin{equation}
    \label{eq:grad_sigma_wd}
    \begin{array}{c}
        \frac{du}{d\mathbf{h}} = \frac{e^{-\mathbf{W}_d\cdot \mathbf{h}}}{(1 + e^{-\mathbf{W}_d\cdot \mathbf{h}})^2} \cdot \mathbf{W}_d^T \\
        \frac{du}{d\mathbf{W}_d} = \frac{e^{-\mathbf{W}_d\cdot \mathbf{h}}}{(1 + e^{-\mathbf{W}_d\cdot \mathbf{h}})^2} \cdot \mathbf{h}^T
    \end{array}
\end{equation}
%

Because the input of \textbf{Encoding Layer} includes target KPI, which is always a constant.
The gradients related to the input KPI become irrelevant. 
Thus we define another approximation referred to the Hebbian Learning Rule to calculate the gradients of $\frac{d\mathbf{h}}{d\mathbf{W}_e}$ in Eq. \ref{eq:dhdw1}.
The equation connects nodes of input, output, and hidden as well as the last time period of these nodes.
\begin{equation}
    \label{eq:dhdw1}
    \frac{d\mathbf{h}}{d\mathbf{W}_e} \approx \left[z, x, \mathbf{v}\right] \cdot \sum_{k}{sign((\mathbf{h}_{k}^{(t)} - \mathbf{h}_{k}^{(t-1)}) \cdot (u^{(t)} - u^{(t-1)}))}
\end{equation}

To avoid outlier disturbing, we use normalized last $\tau$-period average gradients for the backward propagation.
In practice, the last $\tau$ period gradients are logged in $\mathcal{G}_e = \{\frac{dJ}{d\mathbf{W}_e}\}$ and $\mathcal{G}_d = \{\frac{dJ}{d\mathbf{W}_d}\}$. 
The normalized average gradients are calculated in Eq.\ref{eq:avg_grad} and Eq.\ref{eq:norm_grad}.
A learning rate $\eta$ is applied to update weights in Eq.\ref{eq:update_w1}.
In practice, we recommend the learning rate is $0.01$.
Too small may lead to latency, and too large may lead to unstable adjustment values.
\begin{equation}
    \label{eq:avg_grad}
    \nabla{\mathbf{W}_e} = \frac{1}{\tau}\sum^{t+\tau}_{t}\mathcal{G}_e,
    \quad \nabla{\mathbf{W}_d} = \frac{1}{\tau}\sum^{t+\tau}_{t}\mathcal{G}_d
\end{equation}
\begin{equation}
    \label{eq:norm_grad}
    \nabla{\mathbf{W}_e}' = \frac{\nabla{\mathbf{W}_e}}{norm(\nabla{\mathbf{W}_e})},
    \quad\nabla{\mathbf{W}_d}' = \frac{\nabla{\mathbf{W}_d}}{norm(\nabla{\mathbf{W}_d})}
\end{equation}
\begin{equation}
    \label{eq:update_w1}
        \mathbf{W}_e = \mathbf{W}_e - \eta\nabla{\mathbf{W}_e}', \quad
        \mathbf{W}_d = \mathbf{W}_d - \eta\nabla{\mathbf{W}_d}'
\end{equation}
\section{Experiments}
In this section, we conduct experiments on the famous open dataset iPinYou \cite{liao2014ipinyou} and on our production RTB system.
The iPinYou Dataset records three seasons' 24 days' logs and includes 78M bid records with 24M impressions, 20K clicks, and 1K conversions in total.
As other algorithms\cite{liu2020effective}, we select the 2nd season data which have enough conversions for our experiments.
\textbf{(1) Our MCMF has a well performance to gain conversion in Section \ref{sec:base_exp}.
(2) The ablation experiments show the effectiveness of the merged feature in Section \ref{sec:ablation}.
(3) Even under the extreme sparsity feedback, the MCMF can get the best performance in Section \ref{sec:sparse_exp}.
(4) The MCMF is applied on our RTB production and get a well performance in conversion in Section \ref{sec:ab_test}.}
\subsection{Base Experiments}
\label{sec:base_exp}
According to Eq.\ref{eq:bid_adjust_value}, we choose the data of IPinyou first six days for training models to get the $pCTR$ and $pCVR$, and use these day's average PPC as the expected PPC ($PPC_e$).
Comparing methods are \textbf{Non-linear}(\textbf{NL})\cite{liu2020effective}, \textbf{PID} \cite{perlich2012bid}, and \textbf{RL} \cite{cai2017real}.
The MCMF extra input features are the accumulated $CTR$, $CVR$, $pCTR$, and $pCVR$, while the base input only includes the target KPIs and the corresponding feedback.
For the fairness of experiments, conversion is chosen as the bidder's utility for comparison (the more the better), and PPC should not exceed the expected PPC.

The base experiments mainly test all methods' efficiency under different budget and constraint conditions.
We set two budget conditions, \textbf{Adequate} and \textbf{Tight}, and cross them with two types of common constraints, \textbf{Single} and \textbf{Multi}, for a total of four base experiment conditions.
The two kinds of budget is determined according to the average of last 6 days cost.
Actually, the original budget (average cost of last 6 days) is surplus.
Thus, we define an \textbf{Adequate} budget as $1/256$ of the original budget, and a \textbf{Tight} budget as $1/1024$ of the original budget.
We set PPC as the \textbf{Single} constraint and set both PPC and budget as \textbf{Multi}-constraints.
Once the cost exceeds the budget, the bid is terminated.
The results of the experiments are displayed in Table \ref{tab:results}.
The variance in cumulative \textbf{Conversion} and \textbf{Cost} is displayed in Figure \ref{fig:conv_ppc}.
\begin{table}[th]
    \footnotesize
    \vspace{-0.5cm}
    \setlength{\abovedisplayskip}{-0.3cm}
    \setlength{\belowdisplayskip}{-0.5cm}
    \setlength{\abovecaptionskip}{-0.5cm}
    \setlength{\belowcaptionskip}{0cm}
    \centering
    \begin{tabular}{l|cccc|cccc}
        \toprule[1.5pt]
        MD. & IMP & CLK & CONV & PPC & IMP & CLK & CONV & PPC  \\
        \midrule[1.0pt]
        \rule{0pt}{0.5pt}
        & \multicolumn{4}{c}{Adequate} & 
        \multicolumn{4}{c}{Tight} \\ 
        \midrule[0.8pt]
        &\multicolumn{8}{c}{Single Constraint} \\
        \midrule
        NL & 125 & 84 & 17 & 835.29 & 125 & 84 & 17 & 835.29 \\ 
        PID & 276 & 125 & 30 & 861.23 & 230 & 104 & 27 & 845.44 \\
        RL & 762 & 131 & \textbf{33} & 1518.85 & 253 & 21 & 3 & 2053.67 \\
        Ours & 295 & 142 & \textbf{33} & \textbf{786.76} & 248 & 120 & \textbf{29} & \textbf{788.41} \\
        \midrule[0.8pt]
        &\multicolumn{8}{c}{Multi Constraints} \\
        \midrule
        NL & 188 & 110 & 26 & 710.04 & 140 & 84 & 22 & \textbf{579.55} \\
        PID & 221 & 103 & 23 & 856.00 & 177 & 83 & 22 & 636.86 \\
        RL & 534 & 93 & \textbf{27} & 1205.74 & 210 & 17 & 3 & 1724.33 \\
        Ours & 200 & 115 & \textbf{27} & \textbf{676.15} & 201 & 116 & \textbf{26} & 706.81 \\
        \bottomrule[1.5pt]
    \end{tabular}
    \caption{\textbf{Adequate}: $budget=182344, PPC_e=1800$; \textbf{Tight}: $budget=22793, PPC_e=1800$. \textbf{CONV}: the more the better; \textbf{PPC}: not exceeding expected PPC.}
    \label{tab:results}
\end{table}
\begin{figure}[h]
    \vspace{0.1cm}
    \setlength{\abovedisplayskip}{-0.cm}
    \setlength{\belowdisplayskip}{-0.cm}
    \setlength{\abovecaptionskip}{-0.cm}
    \setlength{\belowcaptionskip}{-0.5cm}
    \centering
    \includegraphics[width=0.85\linewidth]{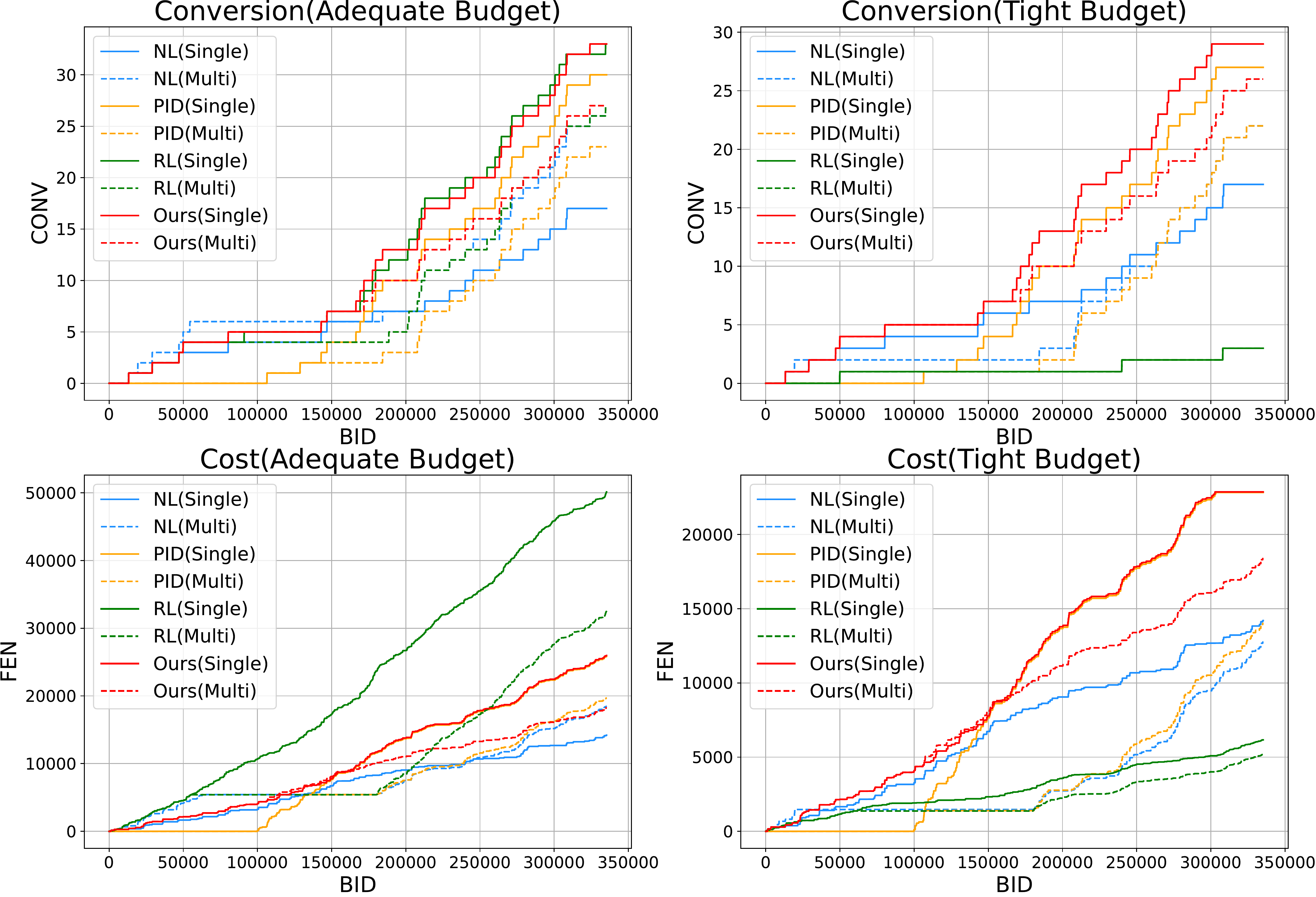}
    \caption{1st Row: \textbf{Conversion}; 2nd Row: \textbf{Cost}. 1st Col.:  \textbf{Adequate}; 2nd Col.: \textbf{Tight}. Solid line: \textbf{Single}; Dot line: \textbf{Multi}.}
    \label{fig:conv_ppc}
\end{figure}

According to the results, for all conditions, our \textbf{MCMF} performs the best in \textbf{Conversion}, and the \textbf{PPCs} are all satisfied with the expected PPC. 
Comparing to the \textbf{RL}, our \textbf{MCMF} get an equal \textbf{Conversions} with a lower cost under the \textbf{Adequate} budget.
Under the \textbf{Tight} budget, the RL's performance of \textbf{Conversion} is far behind ours.
For the \textbf{PID}, the method's \textbf{Conversion} and \textbf{PPC} get average performances.
According to Figure \ref{fig:conv_ppc}, its \textbf{Conversion} and \textbf{Cost} have a noticeable delay in the beginning.
Compared to the \textbf{Single} and \textbf{Multi}-constraint conditions, all methods have a reduction of \textbf{PPC} under the same budget limitation.
However, our \textbf{MCMF} has the capability to reduce the \textbf{PPC} without additional budget pacing or management.
\subsection{Ablation Experiments}
\label{sec:ablation}
We set an ablation experiments to show the contribution of merged features.
The extra input features are divided into posterior features ($CTR$ and $CVR$) and prior features ($pCTR$ and $pCVR$).
Four kinds of input features are tested, including extra features not given (\textbf{NG}), posterior features (\textbf{PO}), prior features (\textbf{PI}) and all of the extra features (\textbf{FULL}).
The results in Table \ref{tab:ablation} show that a trade-off between \textbf{Conversion} and \textbf{PPC} when all features are enabled (\textbf{FULL}).

\begin{table}[h]
    \footnotesize
    \vspace{-0.2cm}
    \setlength{\abovedisplayskip}{-0.3cm}
    \setlength{\belowdisplayskip}{-0.5cm}
    \setlength{\abovecaptionskip}{-0.5cm}
    \setlength{\belowcaptionskip}{0cm}
    \centering
    \begin{tabular}{l|cccc|cccc}
        \toprule[1.5pt]
        Exp. & IMP & CLK & CONV & PPC & IMP & CLK & CONV & PPC  \\
        \midrule[1.0pt]
        \rule{0pt}{0.5pt}
        & \multicolumn{4}{c}{Adequate} & 
        \multicolumn{4}{c}{Tight} \\ 
        \midrule[0.8pt]
        \textbf{NG} & 194 & 116 & 26 & \textbf{697.50} & 194 & 116 & 26 & \textbf{697.50} \\
        \textbf{PO} & 194 & 116 & 26 & \textbf{697.50} & 241 & 118 & 27 & 846.59 \\
        \textbf{PI} & 298 & 142 & \textbf{33} & 798.85 & 245 & 117 & 27 & 844.89 \\
        \textbf{FULL} & 295 & 142 & \textbf{33} & 786.76 & 248 & 120 & \textbf{29} & 788.41 \\
        \bottomrule[1.5pt]
    \end{tabular}
    \caption{\textbf{NG}: Extra features not given. \textbf{PO}: Use posterior features. \textbf{FT2}: Use prior features. \textbf{FULL}: Use all of extra features.}
    \label{tab:ablation}
\end{table}
\subsection{Sparsity Experiments}
\label{sec:sparse_exp}
\begin{figure}[h]
    \vspace{-0.5cm}
    \setlength{\abovedisplayskip}{-0.2cm}
    \setlength{\belowdisplayskip}{-0.cm}
    \setlength{\abovecaptionskip}{-0.cm}
    \setlength{\belowcaptionskip}{-0.5cm}
    \centering
    \includegraphics[width=0.85\linewidth]{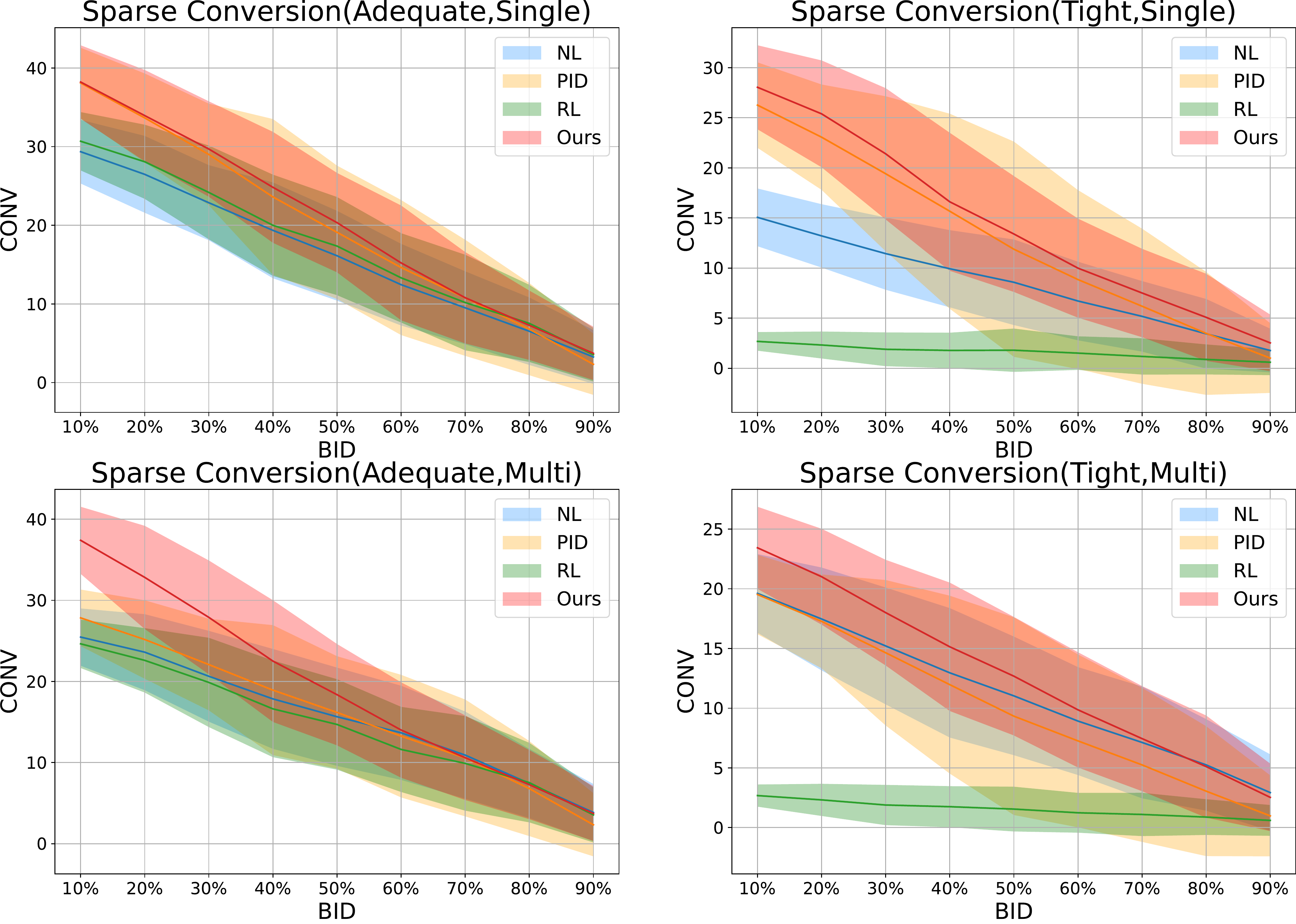}
    \caption{1st Row: \textbf{Single}; 2nd Row: \textbf{Multi}. 1st Col.: \textbf{Adequate}; 2nd Col.: \textbf{Tight}}
    \label{fig:sparse_conv}
\end{figure}

The experiments are designed to demonstrate the performance of all methods under various sparsity circumstances.
We drop out 10\% to 90\% of bids randomly, and all methods attend to the same bid individually.
We perform each experiment 100 times and show the average \textbf{Conversion} with the confidence interval in Figure \ref{fig:sparse_conv}.

Our \textbf{MCMF} get the best performance in most conditions.
On the extreme sparse condition (\textbf{Tight} budget with 90\% of bids dropped out), the \textbf{Conversion} of \textbf{NL} surpasses ours.
We found out that the \textbf{NL} has 2.93 \textbf{Conversion} with \textbf{PPC} of 4130.62 FEN, and our \textbf{MCMF} has 2.53 \textbf{Conversion} with \textbf{PPC} of 2861.41 FEN (the others \textbf{Conversion} less than 1).
That is to say, we receive a comparable average number of \textbf{Conversions} at a 30\% cheaper cost.
The \textbf{PID} has a wider confidence interval, especially in \textbf{Tight} budget.
It indicates that the \textbf{PID} controller is always with uncertainty and instability.
The \textbf{RL} has a comparable performance under \textbf{Adequate} budget, but the worse performance under \textbf{Tight} budget.
\subsection{Online A/B Test}
\label{sec:ab_test}
\begin{figure}[h]
    \vspace{-0.3cm}
    \setlength{\abovedisplayskip}{-0.3cm}
    \setlength{\belowdisplayskip}{-0.cm}
    \setlength{\abovecaptionskip}{-0.cm}
    \setlength{\belowcaptionskip}{-0.5cm}
    \centering
    \includegraphics[width=0.95\linewidth]{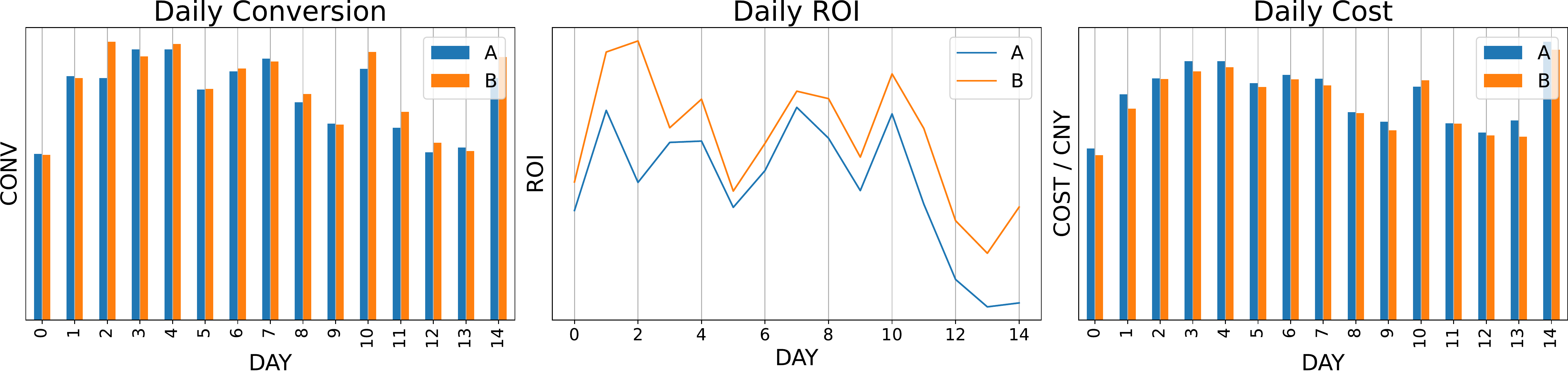}
    \caption{LEFT: Conversion(The more the better); MIDDLE: ROI (The more the better.); RIGHT: Cost}
    \label{fig:abtest}
\end{figure}

We set an A/B test in our online RTB production.
\textbf{A} is the current online approach, and \textbf{B} is our MCMF.
Our advertising includes more than 7K goods, 22K advertisers, and 4M bids with daily ten billion requests.
All of the bidders choose the \textbf{Conversion} as the utility, and use our MCMF to participate in the auction.
The daily performance is illustrated in Figure \ref{fig:abtest}.
For 15 days, the total conversion of our MCMF is 2.96\% more and ROI is 5.55\% greater with 2.46\% fewer expenditures.
\section{Conclusion}
We suggested a bid optimization approach for the Real-Time Bidding called Multi-Constraints with Merging Features (MCMF).
It is a two-layer MLP, with our developed cost function updating the weights using approximated gradients.
Its input incorporates numerous bidding statuses as merging features, which can guarantee the performance under sparse and delayed feedback conditions.
The cost function is proposed to make utility maximization and budget management more consistent.
The approximated gradients are proposed by following the Hebbian Learning Rule, so that the updating can still work in backward propagation even in the absence of bidding environment modeling.
Experiments on a popular open dataset and our RTB production showed that our MCMF achieves more conversions and fewer costs with stable budget management.
\bibliographystyle{unsrt}  
\bibliography{references.bib}

\begin{thebibliography}{10}

\bibitem{yuan2014survey}
Yong Yuan, Feiyue Wang, Juanjuan Li, and Rui Qin.
\newblock A survey on real time bidding advertising.
\newblock In {\em Proceedings of 2014 IEEE International Conference on Service
  Operations and Logistics, and Informatics}, pages 418--423. IEEE, 2014.

\bibitem{kitts2017ad}
Brendan Kitts, Michael Krishnan, Ishadutta Yadav, Yongbo Zeng, Garrett Badeau,
  Andrew Potter, Sergey Tolkachov, Ethan Thornburg, and Satyanarayana~Reddy
  Janga.
\newblock Ad serving with multiple kpis.
\newblock In {\em Proceedings of the 23rd ACM SIGKDD International Conference
  on Knowledge Discovery and Data Mining}, pages 1853--1861, 2017.

\bibitem{tashman2020dynamic}
Michael Tashman, Jiayi Xie, John Hoffman, Lee Winikor, and Rouzbeh Gerami.
\newblock Dynamic bidding strategies with multivariate feedback control for
  multiple goals in display advertising.
\newblock {\em arXiv preprint arXiv:2007.00426}, 2020.

\bibitem{xu2015smart}
Jian Xu, Kuang-chih Lee, Wentong Li, Hang Qi, and Quan Lu.
\newblock Smart pacing for effective online ad campaign optimization.
\newblock In {\em Proceedings of the 21th ACM SIGKDD International Conference
  on Knowledge Discovery and Data Mining}, pages 2217--2226, 2015.

\bibitem{lin2020budget}
Chi-Chun Lin, Kun-Ta Chuang, Wush Chi-Hsuan Wu, and Ming-Syan Chen.
\newblock Budget-constrained real-time bidding optimization: Multiple
  predictors make it better.
\newblock {\em ACM Transactions on Knowledge Discovery from Data (TKDD)},
  14(2):1--27, 2020.

\bibitem{liu2020effective}
Mengjuan Liu, Wei Yue, Lizhou Qiu, and Jiaxing Li.
\newblock An effective budget management framework for real-time bidding in
  online advertising.
\newblock {\em IEEE Access}, 8:131107--131118, 2020.

\bibitem{perlich2012bid}
Claudia Perlich, Brian Dalessandro, Rod Hook, Ori Stitelman, Troy Raeder, and
  Foster Provost.
\newblock Bid optimizing and inventory scoring in targeted online advertising.
\newblock In {\em Proceedings of the 18th ACM SIGKDD international conference
  on Knowledge discovery and data mining}, pages 804--812, 2012.

\bibitem{yang2020motiac}
Chaoqi Yang, Junwei Lu, Xiaofeng Gao, Haishan Liu, Qiong Chen, Gongshen Liu,
  and Guihai Chen.
\newblock Motiac: Multi-objective actor-critics for real-time bidding.
\newblock {\em arXiv preprint arXiv:2002.07408}, 2020.

\bibitem{lu2019reinforcement}
Junwei Lu, Chaoqi Yang, Xiaofeng Gao, Liubin Wang, Changcheng Li, and Guihai
  Chen.
\newblock Reinforcement learning with sequential information clustering in
  real-time bidding.
\newblock In {\em Proceedings of the 28th ACM International Conference on
  Information and Knowledge Management}, pages 1633--1641, 2019.

\bibitem{cai2017real}
Han Cai, Kan Ren, Weinan Zhang, Kleanthis Malialis, Jun Wang, Yong Yu, and
  Defeng Guo.
\newblock Real-time bidding by reinforcement learning in display advertising.
\newblock In {\em Proceedings of the Tenth ACM International Conference on Web
  Search and Data Mining}, pages 661--670, 2017.

\bibitem{he2019identification}
Hao He and Niklas Karlsson.
\newblock Identification of seasonality in internet traffic to support control
  of online advertising.
\newblock In {\em 2019 American Control Conference (ACC)}, pages 3835--3840.
  IEEE, 2019.

\bibitem{karlsson2020feedback}
Niklas Karlsson.
\newblock Feedback control in programmatic advertising: The frontier of
  optimization in real-time bidding.
\newblock {\em IEEE Control Systems Magazine}, 40(5):40--77, 2020.

\bibitem{yang2019bid}
Xun Yang, Yasong Li, Hao Wang, Di~Wu, Qing Tan, Jian Xu, and Kun Gai.
\newblock Bid optimization by multivariable control in display advertising.
\newblock In {\em Proceedings of the 25th ACM SIGKDD International Conference
  on Knowledge Discovery \& Data Mining}, pages 1966--1974, 2019.

\bibitem{munakata2004hebbian}
Yuko Munakata and Jason Pfaffly.
\newblock Hebbian learning and development.
\newblock {\em Developmental science}, 7(2):141--148, 2004.

\bibitem{liao2014ipinyou}
Hairen Liao, Lingxiao Peng, Zhenchuan Liu, and Xuehua Shen.
\newblock ipinyou global rtb bidding algorithm competition dataset.
\newblock In {\em Proceedings of the Eighth International Workshop on Data
  Mining for Online Advertising}, pages 1--6, 2014.

\end{thebibliography}
\end{document}